\documentclass[10pt]{article}
\usepackage{spconf,amsmath,graphicx,multirow,array,url,afterpage,tablefootnote}
\newcolumntype{M}[1]{>{\centering\arraybackslash}m{#1}}

\title{Sound Demixing Challenge 2023 Music Demixing Track Technical Report: TFC-TDF-UNet v3}
\name{Minseok Kim, \ Jun Hyung Lee, \ Soonyoung Jung}
\address{Department of Computer Science, Korea University}

\begin{document}
%
\maketitle
\begin{abstract}
In this report, we present our award-winning solutions for the Music Demixing Track of Sound Demixing Challenge 2023.
First, we propose TFC-TDF-UNet v3, a time-efficient music source separation model that achieves state-of-the-art results on the MUSDB benchmark. We then give full details regarding our solutions for each Leaderboard, including a loss masking approach for noise-robust training. Code for reproducing model training and final submissions is available at \url{github.com/kuielab/sdx23}.

\end{abstract}
\section{Introduction}
\label{sec:intro}
This is a technical report for our solutions for the Music Demixing Track of Sound Demixing Challenge 2023\footnotemark[1] (MDX23). In addition to the standard music source separation (MSS) task conducted in Music Demixing Challenge 2021\cite{mdx21} (MDX21), MDX23 introduced additional challenges: robustness to label-noise and bleeding. 

\footnotetext[1]{www.aicrowd.com/challenges/sound-demixing-challenge-2023}

Label-noise and bleeding are frequently encountered issues in music. 
Label-noise occurs from erratic instrument groupings during automatic metadata-based stem generation in music production. 
Bleeding takes place during music recording sessions when unintended sounds overlap with other instruments. 
This poses a challenge for training source separation models since sources (stem files) may contain instruments that do not belong to the particular class, which requires models to be robust to these errors at training time. Our goal was to enhance the quality of music source separation by addressing these challenges that arise from label-noise and bleeding, in addition to the standard MSS task.

Challenge submissions are ranked into three categories: Leaderboard A for robustness to label-noise, Leaderboard B for bleeding, and Leaderboard C for standard MSS. Furthermore, Leaderboards A and B restrict models to be trained only on specific datasets provided by Moises (namely SDXDB23\_labelnoise and SDXDB23\_bleeding), whereas Leaderboard C does not pose any limitation on training data. 

We first introduce TFC-TDF-UNet v3, the base model architecture for all submissions. Then we describe our approach for each Leaderboard. 
For all experimental results, we use Source-to-Distortion Ratio as the evaluation metric. Throughout the report, ``SDR" will refer to the version used in MDX23 while the other definition of SDR\cite{sdr} will be referred to as ``cSDR" (chunk-level SDR).

\section{TFC-TDF-UNet v3}
\label{sec:v3}
For MDX23 we build upon  TFC-TDF-UNet v2, the spectro-\newline gram-based component of KUIELab-MDX-Net\cite{mdxnet} (award-winning model of MDX21). Our current version, TFC-TDF-UNet v3, achieved top ranks in all Leaderboards. 

\subsection{Improvements}
\label{ssec:v3_improvements}
Here we provide a list of changes that were made to the model structure of TFC-TDF-UNet v2. Our goal was to improve SDR without gaining too much inference time, taking into account the time limit of the MDX23 evaluation system. Changes to v2 that are not listed here (which can be found in our submission code) had negligible effects on performance.

\begin{itemize}
  \item Change overall structure to a ResUnet\cite{ResUnet}-like structure and add a TDF block\cite{mdxnet,tdf} to each Residual Block.
  \item Use Channel-wise Sub-bands\cite{cws} together with larger frequency dimensions.
  \item Train one multi-target model instead of training a single-target model for each instrument class.
  \item Use Instance Normalization and GELU instead of Batch Normalization and ReLU.
  \item Use waveform L2 loss instead of waveform L1.
  \item Add an ``input skip-connection"; concatenate the input spectrogram right before the final convolution (this was effective for multi-source models).
\end{itemize}

Finally, for each Leaderboard, we selected the optimal model hyperparameters using evaluation results on the challenge public set. The final configurations are in Table \ref{tab:cfg}.

\begingroup
\setlength{\tabcolsep}{5pt}
\renewcommand{\arraystretch}{1.0}
\begin{table*}[t]
\centering
\begin{tabular}{l|cc|cc|cc|cc|cc|c}
\hline\hline
\multirow{2}{*}{\centering Model} & \multicolumn{2}{c|}{vocals} & \multicolumn{2}{c|}{drums} & \multicolumn{2}{c|}{bass} & \multicolumn{2}{c|}{other} & \multicolumn{2}{c|}{mean} & \multirow{2}{*}{\centering Speed}\\
\cline{2-11}
 & SDR  & cSDR & SDR  & cSDR & SDR  & cSDR & SDR  & cSDR & SDR  & cSDR \\
\hline
CWS-PResUNet\cite{cwsunet} 
& - & 8.92 & - & 6.38 & - & 5.93 & - & 5.84 & - & 6.77 & - \\ 
KUIELab-MDX-Net\cite{mdxnet} 
& 9.05 & 8.97 & 7.85 & 7.20 & 7.12 & 7.83 & 5.78 & 5.90 & 7.45 & 7.47 & 8.5x \\
Hybrid Demucs\cite{hdemucs} 
& 8.11 & 8.13 & 8.87 & 8.24 & \textbf{7.76} & \textbf{8.76} & 5.39 & 5.59 & 7.53 & 7.68 & 8.9x \\
BSRNN\cite{bsrnn} 
& \textbf{10.04} & \textbf{10.01} & 8.92 & \textbf{9.01} & 6.8 & 7.22 & 6.01 & 6.70 & 7.94 & 8.24 & 0.7x \\
\hline
TFC-TDF-UNet v2\cite{mdxnet} 
& 8.96 & 9.05 & 6.87 & 6.40 & 6.85 & 7.61 & 5.44 & 5.70 & 7.03 & 7.19 & 12.8x \\
TFC-TDF-UNet v3 
& 9.22 & 9.38 & 8.81 & 8.01 & 7.36 & 8.28 & 6.19 & 6.77 & 7.90 & 8.11 & \textbf{15.0x} \\
+ overlap-add 
& 9.34 & 9.59 & \textbf{8.96} & 8.44 & 7.53 & 8.45 & \textbf{6.32} & \textbf{6.86} & \textbf{8.04} & \textbf{8.34} & 3.9x \\

\hline\hline
\end{tabular}
\caption{Performance of TFC-TDF-UNet v3 on the MUSDB18-HQ benchmark. All models are trained solely on the MUSDB18-HQ train set without extra data. We report mean SDR over the test set as well as median cSDR (as in SiSEC18\cite{sisec18}). 
\textbf{Speed} denotes the relative GPU inference speed with respect to real-time on the MDX23 evaluation server 
(speed for BSRNN was measured with an unofficial implementation\protect\footnotemark[2]).}
\label{tab:sota}
\end{table*}
\endgroup

\afterpage{\footnotetext[2]{github.com/crlandsc/Music-Demixing-with-Band-Split-RNN}}

\subsection{Evaluation}
\label{ssec:v3_performance}

For a quantitative comparison with v2 as well as state-of-the-art models, we report performance of TFC-TDF-UNet v3 on the MUSDB18-HQ\cite{musdbhq} (Table \ref{tab:sota}). 
We trained an additional model for MUSDB with the hyperparameters described in Table \ref{tab:cfg}. 
For data augmentation we applied pitch-shift using Soundstretch\footnotemark[3] (semitones $\in\{-3,-2,-1,0,1,2,3\}$) and randomly mixed sources from different songs (remixing)\cite{blend}. 
The v3 model was trained for 47 epochs (we define ``epoch" as 10k steps with batch size 8), which took 3 days using two RTX 3090. 
For early stopping, we stopped training when SDR did not improve by at least 0.05dB within 10 epochs.  

\footnotetext[3]{www.surina.net/soundtouch/soundstretch.html}

For a better comparison with v2, we also report results for v3 without ``overlap-add'' and instead uses the inference method of v2 (trim and concatenate). This made v3 roughly 1.2 times faster than v2 on the MDX23 evaluation server. Even with this lightweight structure, v3 improves v2 by a significant 1.61dB cSDR for ``drums" and 0.92dB on average. Trading off speed for accuracy with overlap-add, TFC-TDF-UNet v3 achieves the highest average SDR/cSDR over all instruments.





\section{Leaderboard A\&B}
\label{sec:ab}
In this section, we present our approach for robust training and details regarding our solutions for Leaderboard A (3rd place) and Leaderboard B (1st place).

\begingroup
\setlength{\tabcolsep}{2.5pt}
\renewcommand{\arraystretch}{1.1}
\begin{table}
\centering
\begin{tabular}{l|ccccc}
\noalign{\smallskip}\noalign{\smallskip}\hline\hline
\multicolumn{1}{c|}{Model} & vocals & drums & bass & other & mean \\
\hline
modelA (label-noise) & \textbf{7.58} & \textbf{6.38} & \textbf{6.43} & \textbf{4.64} & \textbf{6.26} \\
modelA w/o loss masking & 6.12 & 5.31 & 5.31 & 3.45 & 5.05 \\
\hline
modelB (bleeding) & \textbf{7.41} & \textbf{6.20} & \textbf{6.58} & \textbf{4.69} & \textbf{6.22} \\
modelB w/o loss masking & 6.87 & 5.86 & 6.11 & 4.36 & 5.80 \\
\hline
\hline
\end{tabular}
\caption{
Ablation study for loss masking. We report the MDX23 evaluation results. 
The configurations for modelA and modelB follow Table \ref{tab:cfg}.
Note that modelA/modelB are ``single" models and not the final submission ensembles.
}
\label{tab:AB}
\end{table}
\endgroup

\subsection{Data}
\label{ssec:ab_data}
For both Leaderboards, all 203 tracks of the Moises datasets were used for training (SDXDB23\_labelnoise for Leaderboard A and SDXDB23\_bleeding for Leaderboard B). We did not hold out a validation split; doing validation on noisy data did not generalize well. Instead, for early stopping and model selection, we used the challenge public set results where we submitted every 25k steps until mean SDR stopped improving. For data augmentation we used remixing, with no pitch-shift/time-stretch.

\begingroup
\setlength{\tabcolsep}{8pt}
\renewcommand{\arraystretch}{1.0}
\begin{table*}[t]
\centering

\begin{tabular}{l|cc|ccc|c}
\hline\hline
\multicolumn{1}{l|}{} & \multicolumn{2}{c|}{Leaderboard A\&B} & \multicolumn{3}{c|}{Leaderboard C} & \multicolumn{1}{c}{MUSDB}\\ 
Hyperparameter & modelA & modelB & model1 & model2 & model3 & \\

\hline
\textbf{STFT} &  &  &  &  &  &  \\
\multicolumn{1}{l|}{n\_fft} & \multicolumn{2}{c|}{8192} & 8192 & 8192 & 12288 & 8192\\ 
\multicolumn{1}{l|}{hop\_length} & \multicolumn{2}{c|}{1024} & \multicolumn{3}{c|}{2048} & 2048\\

\hline
\textbf{Model} &  &  &  &  &  &  \\
\multicolumn{1}{l|}{\# frequency bins} & \multicolumn{2}{c|}{4096} & \multicolumn{3}{c|}{4096} & 4096\\ 
\multicolumn{1}{l|}{\# inital channels} & \multicolumn{2}{c|}{64} & 128 & 256 & 128 & 160\\ 
\multicolumn{1}{l|}{growth} & \multicolumn{2}{c|}{64} & \multicolumn{3}{c|}{64} & 80\\ 
\multicolumn{1}{l|}{\# down/up scales} & \multicolumn{2}{c|}{5} & \multicolumn{3}{c|}{5} & 5\\ 
\multicolumn{1}{l|}{\# blocks per scale} & \multicolumn{2}{c|}{2} & \multicolumn{3}{c|}{2} & 2\\ 
\multicolumn{1}{l|}{\# sub-bands} & \multicolumn{2}{c|}{4} & \multicolumn{3}{c|}{4} & 4\\
\multicolumn{1}{l|}{TDF b.n. factor\cite{tdf}} & \multicolumn{2}{c|}{4} & \multicolumn{3}{c|}{4} & 4\\
\multicolumn{1}{l|}{normalization} & \multicolumn{2}{c|}{InstanceNorm} & \multicolumn{3}{c|}{InstanceNorm} & InstanceNorm\\ 
\multicolumn{1}{l|}{activation} & \multicolumn{2}{c|}{GELU} & \multicolumn{3}{c|}{GELU} & GELU\\
\multicolumn{1}{l|}{\# parameters} & \multicolumn{2}{c|}{30M} & 46M & 90M & 46M & 70M\\

\hline
\textbf{Training} &  &  &  &  &  &  \\
\multicolumn{1}{l|}{optimizer} & \multicolumn{2}{c|}{Adam} & \multicolumn{3}{c|}{Adam} & Adam\\
\multicolumn{1}{l|}{learning rate} & \multicolumn{2}{c|}{1e-4} & 5e-5 & 3e-5 & 5e-5 & 5e-5\\
\multicolumn{1}{l|}{batch size} & \multicolumn{2}{c|}{6} & \multicolumn{3}{c|}{8} & 8\\
\multicolumn{1}{l|}{chunk size} & \multicolumn{2}{c|}{$\approx6s$} & \multicolumn{3}{c|}{$\approx6s$} & $\approx6s$\\
loss mask dims & batch & batch, time & \multicolumn{3}{c|}{none} & none\\ 
q & $\in[1/3, 1/2)$ & 0.93 & \multicolumn{3}{c|}{n/a} & n/a\\

\hline
\textbf{Inference} &  &  &  &  &  &  \\
\multicolumn{1}{l|}{chunk size} & \multicolumn{2}{c|}{$\approx 24s$} & \multicolumn{3}{c|}{$\approx48s$} & $\approx24s$\\
\multicolumn{1}{l|}{overlap-add factor} & \multicolumn{2}{c|}{8} & \multicolumn{3}{c|}{8} & 4\\ 

\hline
\hline
\end{tabular}
\caption{
Hyperparameter configurations for TFC-TDF-UNet v3 models.  
(\textbf{growth}: the number of channels is increased/decreased by this amount after each down/upsampling layer; 
\textbf{loss mask dims}: the $q$-quantiles are computed along these dimensions for loss masking;
\textbf{overlap-add factor}: hop\_size = chunk\_size / overlap-add\_factor)
}
\label{tab:cfg}
\end{table*}
\endgroup

An important preliminary for Leaderboards A and B was understanding what kind of noise label-noise and bleeding produced. By definition, 1) both corruptions add instrument sounds belonging to other classes and 2) for data with label-noise the loudness of noise would be equal to that of the clean source, while for bleeding the loudness would be lower. For a closer look at how these corruptions were actually simulated, we also listened to several tracks and found that label-noise adds just one instrument belonging to another class, while bleeding seemed to add all other instruments.

\subsection{Noise-Robust Training Loss}
\label{ssec:ab_method}
Since the Moises datasets were corrupted in a way so that manual cleaning would not be possible, 
the main challenge was to design a robust training algorithm for source separation. Our noise-robust training loss, which is basically a loss masking (truncation) method, was clearly effective for this task as shown in Table \ref{tab:AB}. We now describe our method for each Leaderboard.

\subsubsection{Leaderboard A: Label-noise}
\label{sssec:ab_method_a}
If we randomly chunk a noisy target source at training time, each chunk will have different amounts (e.g., duration, loudness) of label-noise. We gain on the fact that some chunks can be clean and these clean chunks can be filtered using its training loss. Intuitively, target source chunks with more noise are likely to produce higher training loss since they lack instrument-related patterns such as timbre. 

To reduce the negative effects of these noisy chunks and train mostly on clean chunks, we use a loss masking scheme where for each training batch, elements with high loss were discarded before weight update. Specifically, for each batch and each class, we masked out per-element losses greater than the $q$-quantile and left $q$ as a hyperparameter. For our final submissions, we used a batch size of 6 and discarded 4 chunks per batch and class.

\begingroup
\setlength{\tabcolsep}{2.5pt}
\renewcommand{\arraystretch}{1.0}
\begin{table*}[t]
\centering
\begin{tabular}{l|l|c|ccccc}
\noalign{\smallskip}\noalign{\smallskip}\hline\hline
\multicolumn{2}{c|}{Model} & \multicolumn{1}{c|}{\multirow{2}{*}{Trainset}} & \multirow{2}{*}{vocals} & \multirow{2}{*}{drums} & \multirow{2}{*}{bass} & \multirow{2}{*}{other} & \multirow{2}{*}{mean} \\
\cline{1-2} \multicolumn{1}{c|}{Architecture} & \multicolumn{1}{c|}{Name} & \multicolumn{1}{c|}{} \\
\hline
Hybrid Demucs & hdemucs\_mmi & \multicolumn{1}{c|}{\multirow{2}{*}{MUSDB trainset (86 songs) + 800 songs}} & 8.82 & 8.77 & 8.93 & 5.97 & 8.13 \\
Hybrid Transformer Demucs & htdemucs\_ft & \multicolumn{1}{c|}{} & 9.02 & 9.19 & 9.56 & 6.23 & 8.51 \\
\cline{1-8}
TFC-TDF-UNet v3 & model1 & \multicolumn{1}{c|}{\multirow{3}{*}{MUSDB (150 songs)}} & 9.44 & 7.79 & 7.73 & 6.16 & 7.79 \\
TFC-TDF-UNet v3 & model2 & \multicolumn{1}{c|}{} & 9.55 & 8.37 & 7.70 & 6.05 & 7.92 \\
TFC-TDF-UNet v3 (vocals only) & model3 & \multicolumn{1}{c|}{} & 9.65 & - & - & - & - \\
\hline
\hline
\end{tabular}
\caption{
Comparison of our Leaderboard C submissions. The rightmost columns show their challenge evaluation results.
}
\label{tab:C}
\end{table*}
\endgroup

\subsubsection{Leaderboard B: Bleeding}
\label{sssec:ab_method_b}
As discussed in Section \ref{ssec:ab_data}, there were more erroneous instruments in bleeding sources than sources with label-noise, which means the amount of noise is more constant throughout the playing time. Consequently, clean random chunks from bleeding data would be rare and harder to obtain. From this inspection, we used a more fine-grained masking scheme where we masked along the temporal dimension as well as the batch dimension. 

But as shown in Table \ref{tab:cfg}, the optimal $q$ value for Leaderboard B models was 0.93, which means only 7\% of the temporal bins were discarded. This may have resulted from the difference in the loudness of noise; compared to label-noise, bleeding was not as harmful and filtering clean chunks was not as important (this can also be inferred from Table \ref{tab:AB} where modelB outperforms modelA when using regular L2 loss).

\subsection{Model}
\label{ssec:ab_model}
For each Leaderboard, the final submission is an ensemble of three TFC-TDF-UNet v3 models trained with the noise-robust training loss of Section \ref{ssec:ab_method}. Each of the three models has the same configurations (following Table \ref{tab:cfg}) but is trained with different random seeds.

\section{Leaderboard C}
\label{sec:c}
We present our approach for the standard MSS task where any training data can be used. Our solution ranked 4th place. 

\subsection{Data}
\label{ssec:c_data}
We used all 150 songs of MUSDB18-HQ for training. As was done for Leaderboards A and B, there was no validation split and submitted every 100k steps instead. For data augmentation we applied pitch-shift (semitones $\in\{-2,-1,0,1,2\}$) and time-stretch (acceleration \% $\in\{-20,-10,0,10,20\}$) as well as remixing. 

\subsection{Method}
\label{ssec:c_method}
The final submission is an ensemble of five models: Hybrid Demucs\cite{hdemucs}, Hybrid Transformer Demucs\cite{htdemucs} and three TFC-TDF-UNet v3 models. For the Demucs models, we used pretrained weights from the official Github repository\footnotemark[4]\footnotetext[4]{github.com/facebookresearch/demucs} ($hdemucs\_mmi$ and $htdemucs\_ft$) each with 2 ``shifts" and 50\% overlap. For the TFC-TDF-UNet v3 models, we used the models specified in Table \ref{tab:cfg}. $model1$ and $model2$ are multi-source v3 models, whereas $model3$ is a single-source model for the ``vocals" class that applies high-frequency truncation\cite{mdxnet}.

SDR performance for each model are shown in Table \ref{tab:C}. Blending\cite{blend} weights were chosen according to these evaluation results.

\bibliographystyle{IEEEbib}
\bibliography{references}

\begin{thebibliography}{10}

\bibitem{mdx21}
Yuki Mitsufuji, Giorgio Fabbro, Stefan Uhlich, Fabian-Robert Stöter, Alexandre
  Défossez, Minseok Kim, Woosung Choi, Chin-Yun Yu, and Kin-Wai Cheuk,
\newblock ``Music demixing challenge 2021,''
\newblock {\em Frontiers in Signal Processing}, vol. 1, 2022.

\bibitem{sdr}
Emmanuel Vincent, Shoko Araki, Fabian Theis, Guido Nolte, Pau Bofill, Hiroshi
  Sawada, Alexey Ozerov, Vikrham Gowreesunker, Dominik Lutter, and Ngoc~QK
  Duong,
\newblock ``The signal separation evaluation campaign (2007--2010):
  Achievements and remaining challenges,''
\newblock {\em Signal Processing}, vol. 92, no. 8, pp. 1928--1936, 2012.

\bibitem{mdxnet}
Minseok Kim, Woosung Choi, Jaehwa Chung, Daewon Lee, and Soonyoung Jung,
\newblock ``{KUIEL}ab-{MDX}-{N}et: A two-stream neural network for music
  demixing,''
\newblock in {\em Proc. the ISMIR 2021 Workshop on Music Source Separation},
  2021.

\bibitem{ResUnet}
Zhengxin Zhang, Qingjie Liu, and Yunhong Wang,
\newblock ``Road extraction by deep residual u-net,''
\newblock {\em {IEEE} Geoscience and Remote Sensing Letters}, vol. 15, no. 5,
  pp. 749--753, may 2018.

\bibitem{tdf}
Woosung Choi, Minseok Kim, Jaehwa Chung, Daewon Lee, and Soonyoung Jung,
\newblock ``Investigating {U}-{N}ets with various intermediate blocks for
  spectrogram-based singing voice separation.,''
\newblock in {\em Proc. International Society for Music Information Retrieval
  Conference~(ISMIR)}, 2020, pp. 192--198.

\bibitem{cws}
Haohe Liu, Lei Xie, Jian Wu, and Geng Yang,
\newblock ``Channel-wise subband input for better voice and accompaniment
  separation on high resolution music,''
\newblock in {\em Proc. Interspeech}, 2020.

\bibitem{cwsunet}
Haohe Liu, Qiuqiang Kong, and Jiafeng Liu,
\newblock ``Cws-presunet: Music source separation with channel-wise subband
  phase-aware resunet,''
\newblock {\em arXiv preprint arXiv:2112.04685}, 2021.

\bibitem{hdemucs}
Alexandre D{\'e}fossez,
\newblock ``Hybrid spectrogram and waveform source separation,''
\newblock in {\em Proc. the ISMIR 2021 Workshop on Music Source Separation},
  2021.

\bibitem{bsrnn}
Yi~Luo and Jianwei Yu,
\newblock ``Music source separation with band-split rnn,''
\newblock {\em arXiv preprint arXiv:2209.15174}, 2022.

\bibitem{sisec18}
Fabian-Robert St{\"o}ter, Antoine Liutkus, and Nobutaka Ito,
\newblock ``The 2018 signal separation evaluation campaign,''
\newblock in {\em Proc. Latent Variable Analysis and Signal
  Separation~(LVA/ICA)}, 2018, pp. 293--305.

\bibitem{musdbhq}
Zafar Rafii, Antoine Liutkus, Fabian-Robert St{\"o}ter, Stylianos~Ioannis
  Mimilakis, and Rachel Bittner,
\newblock ``{MUSDB18-HQ} - an uncompressed version of {MUSDB}18,'' Dec. 2019.

\bibitem{blend}
Stefan Uhlich, Marcello Porcu, Franck Giron, Michael Enenkl, Thomas Kemp, Naoya
  Takahashi, and Yuki Mitsufuji,
\newblock ``Improving music source separation based on deep neural networks
  through data augmentation and network blending,''
\newblock in {\em 2017 IEEE International Conference on Acoustics, Speech and
  Signal Processing (ICASSP)}, 2017, pp. 261--265.

\bibitem{htdemucs}
Simon Rouard, Francisco Massa, and Alexandre D{\'e}fossez,
\newblock ``Hybrid transformers for music source separation,''
\newblock in {\em ICASSP 2023-2023 IEEE International Conference on Acoustics,
  Speech and Signal Processing (ICASSP)}. IEEE, 2023, pp. 1--5.

\end{thebibliography}

\end{document}